\documentclass[aps,nofootinbib,showpacs,twocolumn]{revtex4-1}
\usepackage[CJKbookmarks, pdftex, bookmarksnumbered, bookmarksopen, colorlinks, citecolor=blue, linkcolor=blue]{hyperref}
\usepackage[english]{babel}
\usepackage{amstext,amsmath,amssymb,amsfonts,bbm}
\usepackage[latin1]{inputenc}
\usepackage{graphicx}
\usepackage{epstopdf}
\usepackage{amsthm}
\usepackage{tocvsec2}
\usepackage{enumerate}
\usepackage{subfigure}
\usepackage{color}

\begin{document}

\title{The second law of thermodynamics at the microscopic scale}

\author{Thibaut Josset}
\affiliation{Aix Marseille Univ, Universit\'e de Toulon, CNRS, CPT, Marseille, France}

\date{\today}

\begin{abstract} 
\noindent In quantum statistical mechanics, equilibrium states have been shown to be the typical states for a system that is entangled with its environment, suggesting a possible identification between thermodynamic and von Neumann entropies. In this paper, we investigate how the relaxation toward equilibrium is made possible through interactions that do not lead to significant exchange of energy, and argue for the validity of the second law of thermodynamics at the microscopic scale.
\end{abstract}

\maketitle

	The vast majority of phenomena we witness in every day life are irreversible. From the erosion of cliffs under the repeated onslaughts of the ocean to the erasure of our memories, we experience the unyielding flow of time. Thermodynamics, originally the science of heat engines, allows us to predict simple macroscopic processes such as the melting down of an ice cube forgotten on the kitchen table, by the mean of two laws: the conservation of energy and the growth of entropy.
		
	Among the wealth of things we are given to perceive, there are also a few phenomena showing a high degree of regularity. Through the trajectories of planets and stars in the sky, through the swinging of a pendulum clock, we encounter the immutability of time. These led to the development of mechanics and time-reversible dynamical laws. This description of nature extends below atomic scale provided that quantum variables, represented by non-commuting operators, are introduced.

	The revolution initiated in the second half of the $19^\text{th}$ century by the founders of statistical mechanics was to understand that thermodynamics emerges from the microscopic structure of matter; this idea had actually been suggested by D. Bernoulli in \emph{Hydrodynamica} (1738).
\\

	Recently, a new approach to the foundations of statistical mechanics, sometimes called typicality, has been proposed \cite{Popescu06}. Rather than the usual time or ensemble averages, this alternative viewpoint finds its roots in the genuinely quantum notion of entanglement. For a composite system in a pure state, each component, when taken apart, appears in a probabilistic mixture of quantum states, that is, even individual states can exhibit statistical properties. For instance, in the case of a system in contact with a heat bath, it has been shown that its density matrix is generically very close to a thermal state \cite{Tasaki98, Gemmer03, Goldstein06}. This result has been extended to more complicated couplings between the system and its surrounding (see \cite{Popescu06} for details and proofs). Concretely, one considers a system $\text{S}$ and its environment, subjected to a constraint $\text{R}$ (e.g. fixing the total energy). In terms of Hilbert spaces, that means $\mathcal{H}_\text{R} \subseteq \mathcal{H}_\text{S} \otimes \mathcal{H}_\text{env}$. If the number of degrees of freedom composing the system is small compared to the one of the environment, then for most $|\Psi \rangle \in \mathcal{H}_\text{R}$ (in the sense of the Haar measure on the unit sphere), one gets
	\begin{equation}\label{typicality}
		\hat \rho_\text{S} \equiv \text{Tr}_\text{env} |\Psi \rangle \langle \Psi | \approx \text{Tr}_\text{env} \frac{\mathbb{I}_\text{R}}{\dim \mathcal{H}_\text{R}} .
	\end{equation}
In other words, in the framework of typicality, the actual state of the system at some instant of time is very likely to resemble the equilibrium state. For that reason, it is legitimate to use the latter to estimate standard statistical quantities like expectation values of observables or von Neumann entropy $S_\text{vN} \equiv -k_B\text{Tr}\left( \hat \rho_\text{S} \ln \hat \rho_\text{S} \right)$. 

However, from this perspective, the formulation of a second law of thermodynamics seems at first sight problematic. Indeed, it is usually taken for granted that fine-grained entropy has to be conserved under the microscopic dynamics imposed by Schr\"odinger's equation, and that thermalization may only be read from a growth in a coarse-grained entropy, such as in Boltzmann's $H$-theorem. In this work, we propose a very simple and practical resolution of this issue, based on the possibility of creating entanglement without exchanging energy (cf. \cite{BartschGemmer09, Popescu14} for alternative outlooks on this problem). For clarity, we will focus on a single example, prototype of all irreversible transformations, that is the Joule-Gay-Lussac expansion. The solution suggested here is nonetheless much more general, and is thought to be relevant for all ordinary thermodynamic systems.
\\

\begin{figure}
\includegraphics[width=2.5 in]{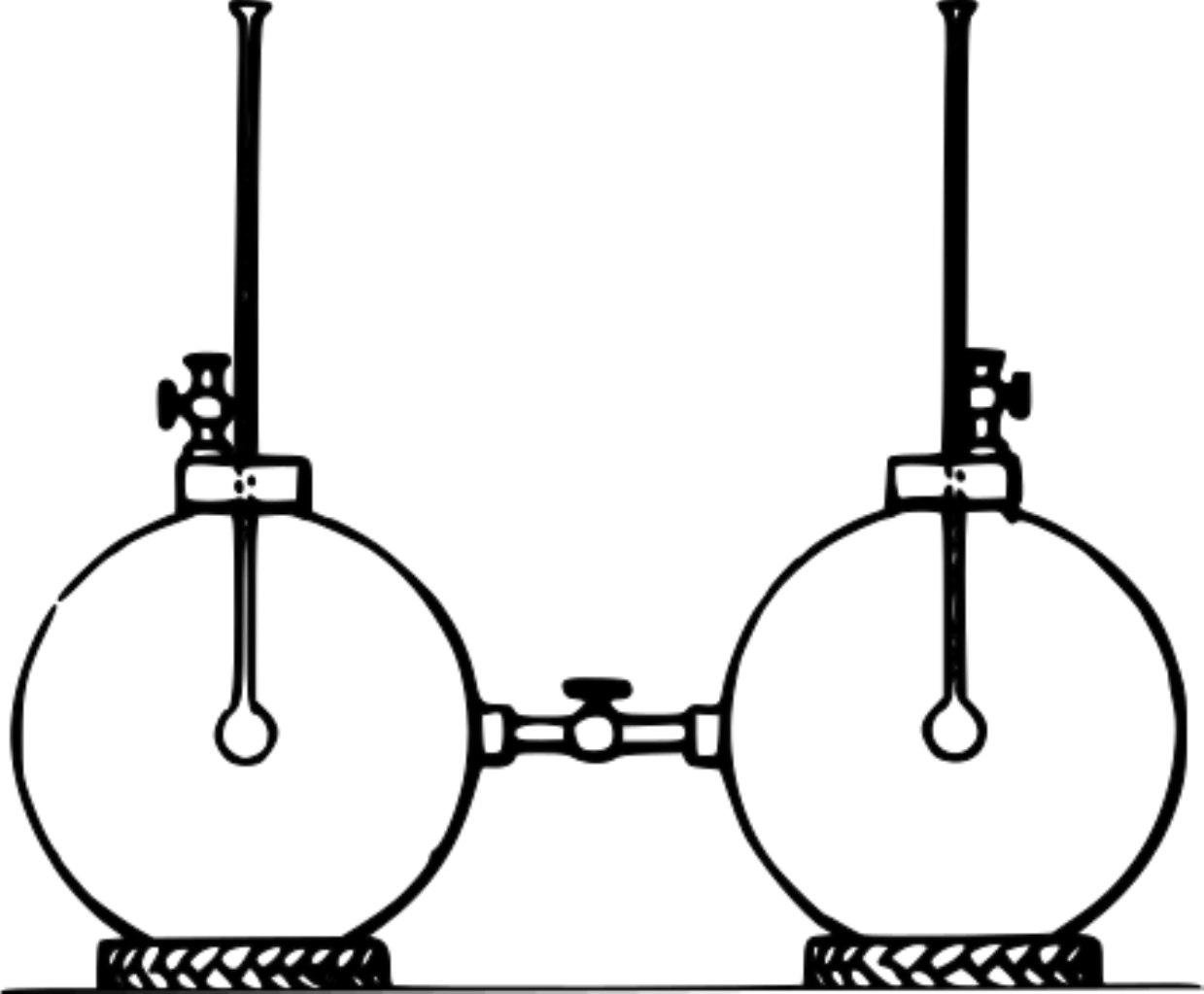}
\caption{Gay-Lussac's experimental setting. Picture taken from G. Lam\'e, \emph{Cours de Physique de  l'\'Ecole Polytechnique} (1836).}
\label{Joule-Gay-Lussac-exp}
\end{figure}

	The celebrated experiment, initially studied by Gay-Lussac and used later by Joule, has played a central role in the history of thermodynamics. It consists in the adiabatic free expansion of a gas, initially kept in one side of a thermally isolated container while the other side is empty (see Fig. \ref{Joule-Gay-Lussac-exp}). When the tap separating the two flasks is opened, the gas starts spreading out until homogeneity is achieved. If the internal energy functional does not depend on the volume, then the initial and final temperatures are found to be equal. The treatment of the Joule-Gay-Lussac expansion for an ideal gas can be found in all textbooks of thermodynamics and is as follows: There is no exchange of heat nor work between the gas and the outside, and, while the accessible volume is doubled, the entropy of the gas, and therefore of the whole Universe, increases by
		\begin{equation} \label{DeltaSth} 
			\Delta S_\text{th} = Nk_B \ln 2 ,
		\end{equation}
indicating the irreversibility of the process.
	
	On the other hand, it is also instructive to have a mechanical description of the Joule-Gay-Lussac expansion. Assuming for simplicity a monoatomic ideal gas, the dynamics of the $N$ particles is governed by the Hamiltonian 
	\begin{equation} \label{Hcl} 
		H = \sum_i \frac{{{\bf p}_i}^2}{2m} + V( {\bf x}_i) ,
	\end{equation}
where  $V$ is the potential keeping the particles inside the whole container. Liouville's theorem then ensures the conservation of the volume-form induced by the symplectic structure, that is to say, there can be no loss of information whatsoever at the microscopic scale. Shifting to quantum mechanics, the evolution is given by the unitary operator
	\begin{equation} \label{Hq} 
		\hat U (t) = e^{-i \hat H t / \hbar} ,
	\end{equation}
where $\hat H$ is simply obtained by canonical quantization of the classical Hamiltonian \eqref{Hcl}. Assuming thermal equilibrium, the initial state for the gas reads
	\begin{equation} 
		\hat \rho_0 = \frac{e^{-\beta \hat H_0}}{\text{Tr}\left(e^{-\beta \hat H_0}\right)} \quad \text{with} ~\hat H_0 = \sum_i \frac{{\hat {\bf p}_i}^2}{2m} + V_0(\hat {\bf x}_i),
	\end{equation}
where $V_0$ is the potential constraining the particles in one flask only, and $\beta$ is the inverse temperature at which the gas has been prepared. While the von Neumann entropy of the gas initially matches with the thermodynamic one, it is inevitably preserved under the unitary evolution \eqref{Hq}, namely
	\begin{equation} \label{DeltaSvn}
		\Delta S_\text{vN} = 0 . 
	\end{equation}

The tension between the irreversible macroscopic thermodynamics \eqref{DeltaSth} and the reversible microscopic mechanics \eqref{DeltaSvn} is manifest. Our proposal is that an increase of von Neumann entropy is actually allowed by a more careful analysis of what a thermally isolated system is.
\\

	Consider a classical particle arriving with a momentum $p$ perpendicularly to a wall initially at rest. Assuming an elastic collision, and given the large ratio between the masses of the wall and the particle, one obtains that after the bounce, the particle has a momentum $-p$, while the wall has acquired a momentum $2 p$. The kinetic energy transferred to the wall being negligible, the energy of the particle is conserved over the bounce with a very good approximation. Consequently, the classical dynamics of the gas is well described by the effective Hamiltonian \eqref{Hcl}.
	
	From the point of view of quantum mechanics, the situation is more delicate. The total energy and momentum are also conserved during a rebound, so that the state for the composite system made of the particle and the wall evolves from $|p \rangle \otimes |0 \rangle$ to $|{-}p\rangle \otimes |2p \rangle$. By linearity, if the particle started in a superposition $|\phi\rangle = \int \phi(p) | p \rangle d p$, then the unitary transformation associated to the bounce reads
	\begin{equation} 
		|\phi\rangle \otimes |0 \rangle \rightarrow  \int \phi( p) |{-} p \rangle \otimes | 2p \rangle dp ,
	 \end{equation}
and, after tracing out the wall's degrees of freedom, the evolution of the state of the particle becomes
	\begin{equation} \label{bounce}
		|\phi \rangle\langle \phi | \rightarrow \int |\phi( p)|^2 |{-} p \rangle \langle{-} p | d p ,
	\end{equation}
which is, in general, not unitary. Hence, the bounce is accompanied with a rise of von Neumann entropy. 

	The model presented above is of course very schematic, the physics of the interaction between the gas and the container, and the initial state of the latter, are in fact much more complicated. However, it already shows that such an interaction must affect their quantum correlations, and can lead to an increase of von Neumann entropy for the gas (see \cite{GemmerMahler01} for a general result). In other words, at the quantum level, the energy distribution of the gas is conserved yet the evolution is not unitary, and the naive canonical quantization of \eqref{Hcl} does not strictly apply.
	
	It is now possible to derive the final state for the gas from its quantum mechanical description alone. The space $\mathcal{H}_\text{R}$ of accessible states is such that the energy distribution for the gas remains unchanged (for instance, $\langle \hat H \rangle = \frac{3}{2} N k_\text{B} T$). After a transient phase, characterized by a relaxation time $\tau$, during which the gas and its environment get more entangled, the state of the Universe $|\Psi(t) \rangle $ can be seen as a generic unit vector in $\mathcal{H}_\text{R}$. Following the conclusion of typicality \eqref{typicality}, we find that the gas reaches a new equilibrium state, namely
	\begin{equation} \hat \rho(t) \equiv \text{Tr}_\text{env}|\Psi(t) \rangle \langle\Psi(t) | \approx  \frac{e^{-\beta \hat H}}{\text{Tr}\left(e^{-\beta \hat H}\right)} \quad (t>\tau). \end{equation}
	As a consequence, after a long-enough time, the variation of von Neumann entropy of the gas matches the increase of entropy expected from thermodynamics.
\\

	Let us now have a quick look at a slightly different experiment: the adiabatic reversible expansion of a gas pushing a piston. Contrary to the Joule-Gay-Lussac expansion, the temperature of the gas is decreasing as work is performed on the piston. During this process, the space $\mathcal{H}_\text{R}$ of accessible states is changing, but its dimension remains constant since the size of the container and the thermal de Broglie wavelength are increasing in the same way. Later, if the piston is slowly pushed back, the gas reverts to its initial temperature. In this transformation, reversibility cannot be understood as the possibility to return to the initial microstate, which is practically impossible as soon as a few particles are involved, but only as conservation of von Neumann entropy. 
	
	On the contrary, when a constraint initially applied to the system is suddenly relaxed, like in Joule-Gay-Lussac's experiment, the space of accessible states is enlarged. The new states, of possibly higher entropy, can only be reached through interaction with the environment. However, this interaction does not need to involve significant energy transfer, the quantum system remains thermally isolated, and the growth of entanglement is responsible for the irreversibility of its evolution.
\\

	Let us mention that the idea of varying entropy without exchanging energy has recently been suggested as a solution to the information loss paradox in black holes evaporation \cite{UnruhWald95,Unruh12,Perez15}. More precisely, at the very last stage of its life, a black hole is believed to have lost almost all its mass in the form of Hawking's radiation, leaving a planckian size object and a matter field in a thermal state, with an entropy proportional to the initial black hole area. Hence, it has been argued that a huge amount of information should be encoded adiabatically into correlations with geometric degrees of freedom, in order to ensure purity of the final quantum state.
\\

	To summarize, the second law of thermodynamics, characterizing irreversible transformations, is usually considered at a mesoscopic level, by the mean of a coarse-grained entropy, while no information loss nor thermalization is expected to occur for the fundamental degrees of freedom. On the other hand, the framework of typicality attempts to make use of properties of quantum systems, specifically entanglement, as foundations for equilibrium statistical mechanics. In particular, it identifies thermodynamic and von Neumann entropies. In that context, it is legitimate to raise the issue of thermalization, namely how an isolated system can reach the maximal entropy state. The observation that, for quantum systems, being thermodynamically isolated is qualitatively very different from being exactly isolated, opens the possibility of nontrivial quantum interactions without significant exchange of energy, and provides a physical ground for the second law at the microscopic scale. It is worth noting that recent work involving heat baths in squeezed states also points out the role of quantum correlations in thermodynamic processes \cite{Rossnagel14}.

	Finally, this approach offers a fresh look at irreversibility. If the entropy of a system evolving unitarily has to be constant, it is not the case for those of its components. Indeed, the nonextensivity of von Neumann entropy allows for subsystems to see their entropy growing, due to a stronger entanglement between them, without any contradiction with the conservation of total information. That might be of relevance for explaining the arrow of time in cosmology: any region of space is constantly interacting with new degrees of freedom entering its past light cone, and, assuming a separable initial state, new quantum correlations can be established. This viewpoint may help us to understand the emergence, locally, of classical mixtures rather than highly-quantum states, and the increase of entropy for the observable universe.

\section*{Acknowledgments}
	
	We are grateful to Goffredo Chirco, Tommaso De Lorenzo, Alejandro Perez, and Carlo Rovelli, for interesting discussions and useful comments.
		

\end{document}